\documentclass{article} 
\usepackage{spconf}
\usepackage{amsmath} 
\usepackage{subfigure}
\usepackage{epsfig}
\usepackage{graphicx}
\usepackage{psfrag}

\begin{document} 

\title{Fast probabilistic snake algorithm}

\name{J{\'e}r{\^o}me Gilles and Bertrand Collin}
\address{DGA$\setminus$Centre Technique d'Arcueil\\ 
16 bis, av. Prieur de la C{\^o}te d'Or, 94114 Arcueil Cedex, France\\ Jerome.Gilles@etca.fr\\
Bertrand.Collin@etca.fr
}
\maketitle

\begin{abstract}
Few people use the probability theory in order to achieve image segmentation with snake models.
In this article\footnote{\textit{International Conference on Image Processing (ICIP), Barcelona, Spain, September 2003}}, we are presenting an active contour algorithm based on a probability approach inspired by A. Blake work and P. R{\'e}fr{\'e}gier's team research in France. Our algorithm, both very fast and highly accurate as far as contour description is concerned, is easily adaptable to any specific application.\\
\\
\textit{Keywords:} snake, energy minimization, probability, regularization.

\end{abstract}
\section{Introduction}

In the last fifteen years the active contours have been successfully applied in many different ways. Snake can be modelled
according to several mathematical formulation. Each one of them entails its own drawback but provides its own advantages at the same time. In 1988, Kass et al. \cite{kass} 
proposed an energy-based formulation,

\begin{equation}
\begin{split}
E(C)=\int_{0}^{1}\alpha(s)\left\| \dfrac{\partial C(s)}{\partial s}\right\|^{2}+\beta(s)\left\| \dfrac{\partial^{2} C(s)}{\partial s^{2}}\right\|^{2}\\
-\left\|\nabla I(C(s))\right\|^{2}ds
\end{split}
\end{equation}

in which $C$ represented the curve, $I$ the image, $\alpha$ and $\beta$ the parameters used for the control of the curve properties. 
Though it works, Kass's model is limited in its sensitivity to initialization. What is more, it has no natural adaptability to changes in topology. As for the improvements proposed (see \cite{cohen},\cite{elomary}), they add to the complexity of the implementation.

In 1997 geodesic active contours were proposed as a geometric alternative to snake (see \cite{sapiro},\cite{kichen}). They included as well an energy-based term. Yet they managed to overcome the deficiencies of the classical snake formulation. Thanks to geodesic active region \cite{paragios} and to levelset implementation, geodesic active contours stopped being sensitive to initialization and at the same time topology changes became implicit.

Lately the question has been examined from an innovative angle by A.~Black et al. \cite{blake}  who resort to probabilities and B-Spline curves. And as for him C.~Chesnaud \cite{chesnaud} proposes a region-based criterion so as to minimize,
\begin{equation}
J(s,\vec{C})=N_{a}(\vec{C})H(\widehat{\theta}_{a})+N_{b}(\vec{C})H(\widehat{\theta}_{b})+K_{l}
\end{equation}
In which $N_i(\vec{C})$ is the number of pixels inside the curve and $H_i(\widehat{\theta_i})$ represents a measure of probability wich depends on the \textit{a priori} model chosen to figure the pixels distribution, see \cite{chesnaud} for more details (we shall call this model \textit{CASP}).\\

\textit{CASP} is actually interesting but its main drawback is that it cannot be used with open curves. That is why we have endeavoured to develop a new formulation of \textit{CASP} so as to adapt it equally to open or closed curves.
The following section gives the formulation of our model, the third one exposes the implementation of our algorithm and the last one states our results and compares them with other models.
\section{Statistical snake model}
Let us assume that:

\begin{itemize}
\item $I$ is the image
\item $C(s)$ represents the curve wich corresponds to $N$ knots ($N_i$)
\item $B$ represents the edges in the image
\item $p(C|B)$ is the probability density that the snake is on the edge
\item $p(B|C)$ is the probability density that there is an edge on the current position of the snake
\end{itemize}

According to the Bayes' rule we can write,
\begin{equation}
p(C|B) = \dfrac{p(B|C)p(C)}{p(B)}
\end{equation}

Since $p(B)$ is constant, the final snake position is the curve $C$ which minimizes the likelihood $p(C|B)$, or\\$p(B|C)p(C)$.\\ If we consider that each knot is independent we can write,

\begin{equation}
p(B|C) = \prod_{i\in[0..N-1]}p(N_i|I)
\end{equation}

The choice of the density $p(N_i|I)$ depends on the problem we want to solve. For example, if we want to detect the edges in the image, we can calculate the variance inside a window centered in $N_i$ (see next section).\\
The density $p(C)$ represents the constraints on the curve such as \cite{chesnaud} where,

\begin{equation}
p(C)=A\exp\{-\frac{1}{2\varphi^2}\sum_{i=0}^{N-1}d_i^2\}
\end{equation}

in which $d_i$ is the distance between $N_i$ and the center of segment $N_{i-1}N_{i+1}$. It limits the irregularity of the curve. Another example of density will be given in the next section.\\
The upper model has a very simple formulation and we can easily adapt it to different problems. A drawback is that the topology changes are not included in the formulation.
\section{Implementation of the statistical snake}
\begin{figure}[t]
\psfrag{ni}{\tiny{$N_i$}}
\psfrag{ni+1}{\tiny{$N_{i+1}$}}
\psfrag{ni-1}{\tiny{$N_{i-1}$}}
\psfrag{l}{\tiny{$L$}}
\psfrag{w}{\tiny{Window}}
\fbox{\includegraphics[width=0.35\linewidth,height=2.5cm]{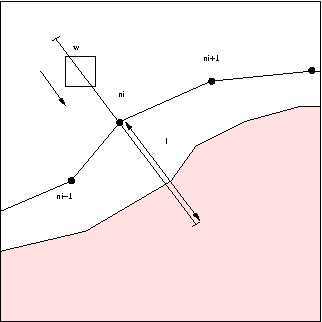}}
\fbox{\includegraphics[width=0.6\linewidth,height=2.5cm]{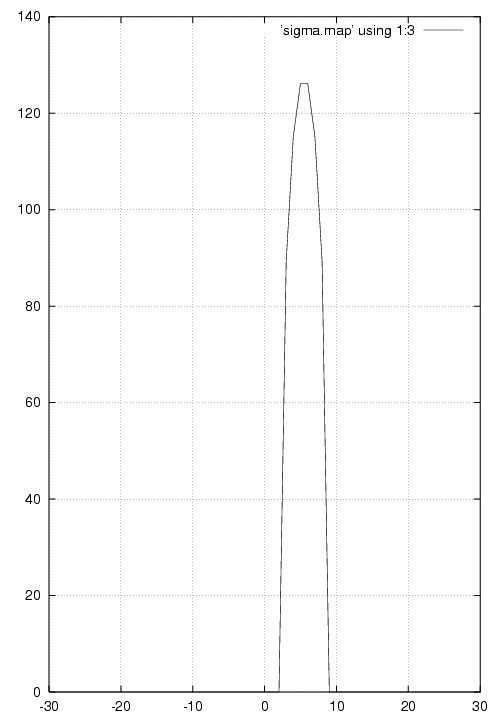}}
\caption{Ideal edge image and variance along the normal of $N_i$}
\label{fig:density}
\end{figure}

In this section, we will give some practical pieces of information to implement our model. As in the case of geodesic snake and, as in \cite{blake}, we consider only the movement along the normal of each knot of the curve. Along this normal, we fix a research depth (named $L$), in order to find the new position of the current knot. On each point of this segment we calculate a measure of variance inside a window so as to estimate $p(N_i|I)$. Thus an edge corresponds to a maximum of the variance (see Fig.\ref{fig:density}).

\begin{figure}[t]
\subfigure[]{\fbox{\includegraphics[width=0.299\columnwidth,height=2.5cm]{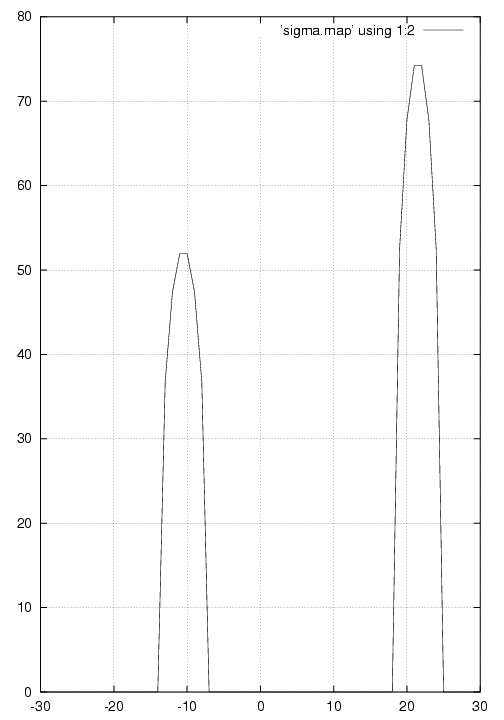}}}
\subfigure[]{\fbox{\includegraphics[width=0.299\columnwidth,height=2.5cm]{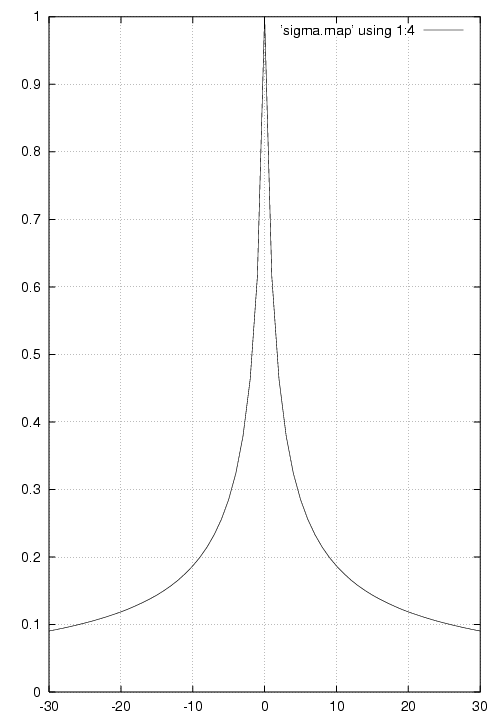}}}
\subfigure[]{\fbox{\includegraphics[width=0.299\columnwidth,height=2.5cm]{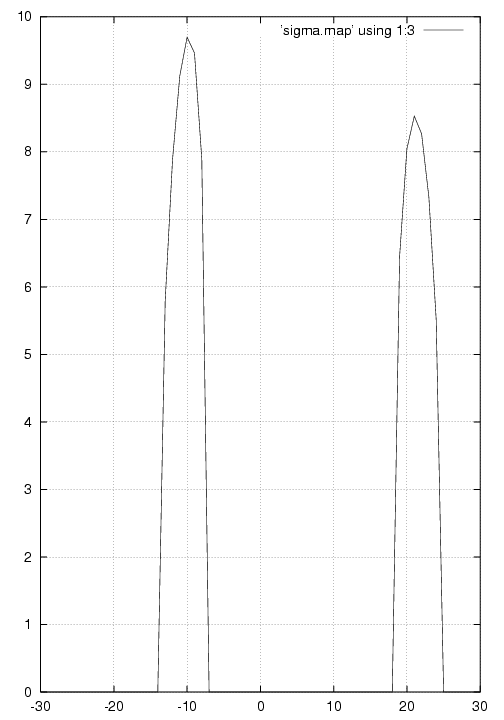}}}
\caption{Regularization: (a) $p(N_i|I)$, (b) the regularity function, (c) $p(N_i|I)$ regularized.}
\label{fig:regul}
\end{figure}

If we intend to regularize the snake, we can use a function which directly acts on the density function $p(N_i|I)$. For example if we want to give top priority to the nearest contour to the current position, we can filter $p(N_i|I)$ with a ``low-pass filter'' considering that ``low frequencies'' corresponds to the positions close to the current knot and ``high frequencies'' to the distant positions (see Fig.\ref{fig:regul}).

With open curves, the normal chosen at end points is orthogonal to the last segment.

So as to improve the convergence of the snake, we process three steps:
\begin{itemize}
\item Optimizing the curve with a few points and high research depth.
\item Resampling the snake to have high resolution contour.
\item Re-optimizing the curve with all points but with a little research depth.
\end{itemize}
\section{Experimental results}
Here we propose to compare first our model with the classical snake model and the \textit{CASP} model of \cite{chesnaud} in the case of closed snakes. Then we will compare our model with the classical one in the case of open snakes.

\subsection{Closed curves}
\begin{figure}[t!]
\centering
\subfigure[]{\includegraphics[width=0.45\columnwidth]{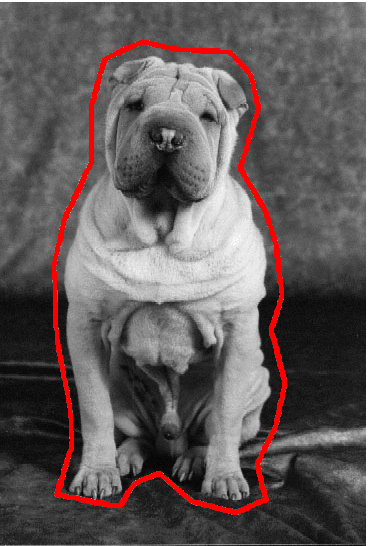}}
\subfigure[]{\includegraphics[width=0.45\columnwidth]{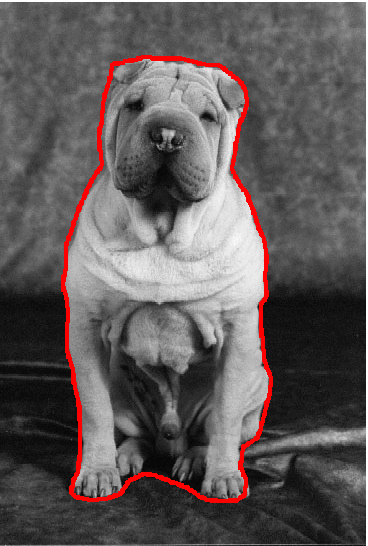}}
\subfigure[]{\includegraphics[width=0.45\columnwidth]{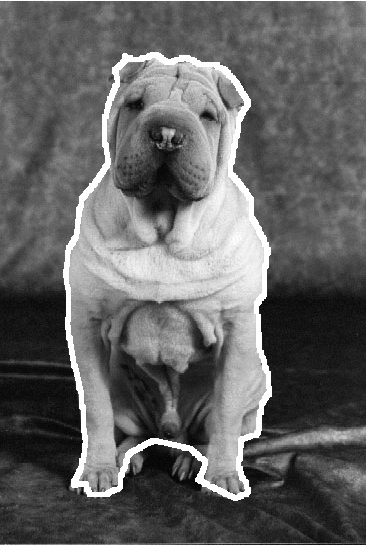}}
\subfigure[]{\includegraphics[width=0.45\columnwidth]{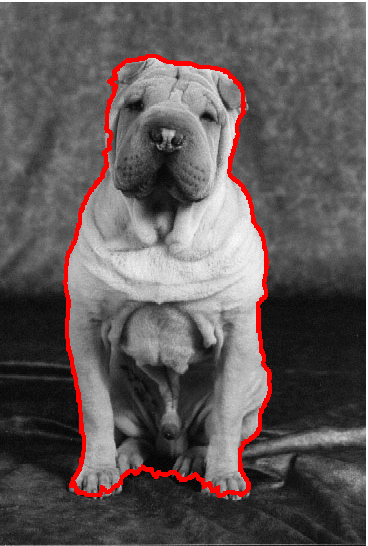}}
\caption{Results in the closed curve case obtain. Initialization (a), classical snake (b), \textit{CASP} model (c) and our statistical model (d)}
\label{fig:closeresult}
\end{figure}

Figure \ref{fig:closeresult} shows the results in the closed curve case. In (a) we have the initialization, in (b) we can see classical snake result with the parameters: $\alpha=0.1$, $\beta=0.1$, $\lambda=0.5$ where $\lambda$ is the regularisation parameter in the evolution equation (see \cite{elomary2}). In (c) we can see the output of the \textit{CASP} algorithm when we choose \textit{max}$\_$\textit{deviation}$=5$, \textit{Nb}$\_$\textit{iteration}$=3000$, regularization$=0.2$ and a gaussian density for each region. The last, (d), shows the result with our model with:

\begin{itemize}
\item First pass: research depth = 25, window analysis size of 7 pixels, regularisation = 0
\item Resampling with a maximum of 4 pixels between two consecutive knots
\item Second pass: research depth = 5, same window size and regularisation = 1
\end{itemize}

Our model gives better resolution than the classical snakes. It is as efficient as \textit{CASP} but our parameters can be found more easily than any.

Tab.\ref{tab:close} gives the different computing times for the two algorithms. The tests were made on a bi-P3-866 with 1Gb of RAM.
We do not give the computing time for the \textit{CASP} model because we haven't implemented the fastest algorithm described in \cite{chesnaud}.

\begin{table}[ht]
\begin{center}
\begin{tabular}{|c|c|}
\hline  Algorithm & Time \\
\hline  Classical snake & 566$\mu$s \\
\hline  Our model & 9.95$\mu$s \\
\hline
\end{tabular}
\end{center}
\label{tab:close}
\caption{Time convergence for closed curves}
\end{table}

\begin{figure}[t!]
\centering
\includegraphics[width=\columnwidth]{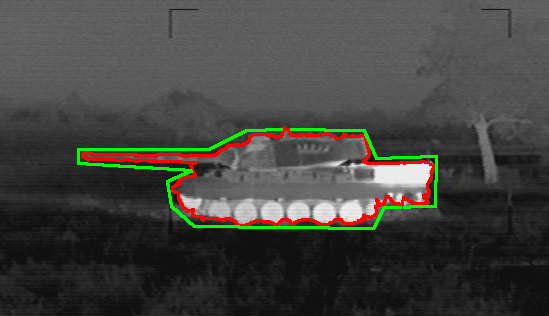}
\caption{Exemple of tank segmentation}
\label{fig:tank}
\end{figure}

Fig.\ref{fig:tank} gives another example of segmentation in infrared image for military application. We use the model with only 14 knots for the first step, resampling with a maximum of 4 pixels between two consecutive knots in the second step and a new optimizing pass in the third step. As we can see, the tank is extracted with high accuracy.

\subsection{Open curves}
\begin{figure}[t!]
\centering
\subfigure[]{\includegraphics[width=0.45\columnwidth]{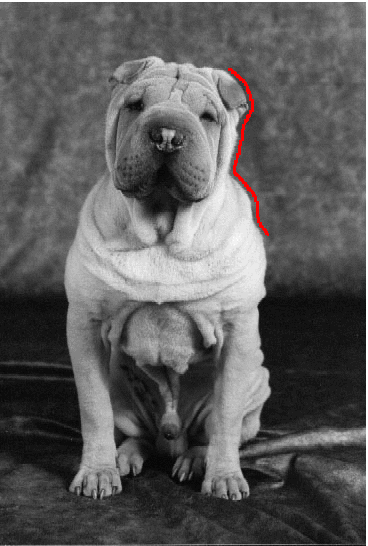}}
\subfigure[]{\includegraphics[width=0.45\columnwidth]{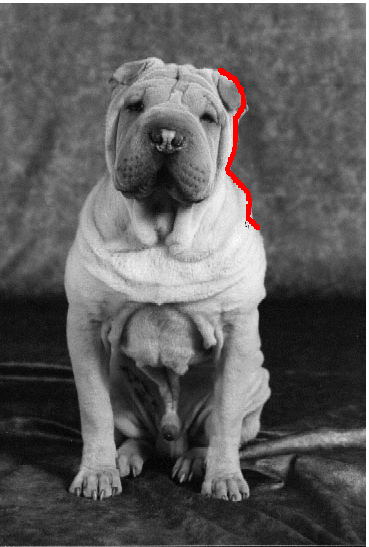}}
\subfigure[]{\includegraphics[width=0.45\columnwidth]{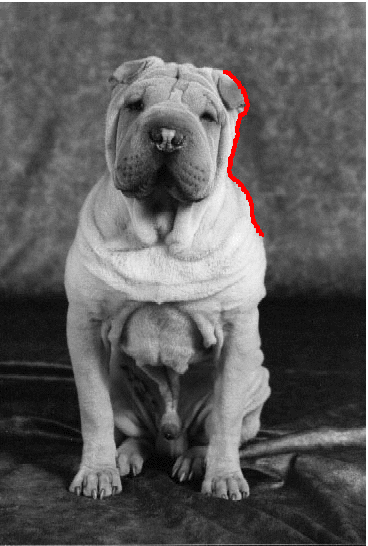}}
\caption{Results in the open curve case obtain. Initialization (a), classical snake (b), our statistical model (c)}
\label{fig:openresult}
\end{figure}

Fig.\ref{fig:openresult} highlights the results obtained on open curves. Initialization is shown in (a). (b) gives the results of the classical snake using $\alpha=0.1$, $\beta=0.5$, $\lambda=0.99$. (c) depicts our model results using the following steps:

\begin{itemize}
\item First pass: research depth = 20, window analysis size of 7 pixels, regularisation = 0
\item Resampling with a maximum of 4 pixels between two consecutives knots
\item Second pass: research depth = 5, same window size and regularisation = 0
\end{itemize}

The parameters that control the classical snake were found with difficulty.
Tab.\ref{tab:open} gives the different computing times showing that our algorithm if 200 times faster than the classical snake model.

\begin{table}[ht]
\begin{center}
\begin{tabular}{|c|c|}
\hline  Algorithm & Time \\
\hline  Classical snake & 565$\mu$s \\
\hline  Our model & 2.2$\mu$s \\
\hline
\end{tabular}
\end{center}
\label{tab:open}
\caption{Time convergence for open curve}
\end{table}

\section{Conclusion}
We have proposed a very fast snake algorithm based on a probability approach. This model accuracy is quite high when using two steps that speed the convergence using few knots and then resample the curve. Futhermore this model uses only three parameters, one of them (the window size) being set to a fixed value for most applications.\\[5mm]
We are now working on its adaptability to fit multiple objects detection (topology changes) and on the regularization process to increase its stability.

\bibliographystyle{IEEEbib}
\bibliography{icip2003}

\begin{thebibliography}{1}

\bibitem{kass}
M.Kass, A.Witkin, and D.Terzopoulos,
\newblock ``Snakes: actives contours models,''
\newblock {\em International Journal of Computer Vision}, vol. 1, pp. 312--331, 1988.

\bibitem{cohen}
L.D.Cohen,
\newblock ``On active contour models and balloons,''
\newblock {\em CVGIP: Image understanding}, vol. 53, no. 2, pp. 211--218, 1991.

\bibitem{elomary}
Y.Elomary and J-M.Chassery,
\newblock ``Cooperation between active contours and multiresolution for edge-based segmentation,''
\newblock {\em Proceedings of the IEEE IMDSP Conference, Cannes, France}, pp. 206--207, 1993.

\bibitem{sapiro}
V.Caselles, R.Kimmel, and G.Sapiro,
\newblock ``Geodesic active contours,''
\newblock {\em International Journal of Computer Vision}, vol. 22, pp. 61--79, 1997.

\bibitem{kichen}
S.Kichenassamy, A.Kumar, P.Olver, A.Tannenbaum, and A.Yezzi,
\newblock ``Gradient flows and geometric active contour models,''
\newblock {\em IEEE International Conference on Computer Vision}, vol. 1, pp. 810--815, 1995.

\bibitem{paragios}
N.K.Paragios,
\newblock {\em Geodesic active regions and level set methods: contributions and applications in artificial vision},
\newblock Ph.D. thesis, University of Nice Sophia Antipolis, France, 2000.

\bibitem{blake}
A.Blake and M.Isard,
\newblock {\em Active Contours},
\newblock Springer, 1999.

\bibitem{chesnaud}
C.Chesnaud,
\newblock {\em Techniques statistiques de segmentation par contour actif et mise en \oe uvre rapide},
\newblock Ph.D. thesis, ENSPM, University of Aix-Marseille, France, 2000.

\bibitem{elomary2}
Y.Elomary,
\newblock {\em Mod\`eles d\'eformables et multir\'esolution pour la d\'etection de contours en traitement d'images},
\newblock Ph.D. thesis, University of Joseph Fourier, Grenoble, France, 1994.

\end{thebibliography}

\end{document}